\documentclass[epj]{webofc}
\usepackage[utf8]{inputenc}
\usepackage[varg]{txfonts}   % Web of Conferences font
\usepackage{booktabs}
\usepackage{xcolor}
\definecolor{darkred}{rgb}{0.4,0.0,0.0}
\definecolor{darkgreen}{rgb}{0.0,0.4,0.0}
\definecolor{darkblue}{rgb}{0.0,0.0,0.4}
\usepackage[bookmarks,linktocpage,colorlinks,
    linkcolor = darkred,
    urlcolor  = darkblue,
    citecolor = darkgreen]{hyperref}
\usepackage{subfigure}
\wocname{EPJ Web of Conferences}
\woctitle{Lattice2017}
\begin{document}
%%%%%%%%%%%%%%%%%%%%%%%%%%%%%%%%%%%%%%%%%%%%%%%%%%%%%%%%%%%%%%%%%%%%%%%%%%%%%
%
\selectlanguage{english}
%----------------------------------------------------------------------------
\title{%
Lattice QCD Application Development within the US DOE Exascale Computing Project
}
%----------------------------------------------------------------------------
\author{%
\firstname{Richard} \lastname{Brower}\inst{1} 
 \and
\firstname{Norman}  \lastname{Christ}\inst{2}
 \and
\firstname{Carleton} \lastname{DeTar}\inst{3}\fnsep\thanks{Speaker, \email{detar@physics.utah.edu}} 
 \and
\firstname{Robert}  \lastname{Edwards}\inst{4}
 \and
\firstname{Paul} \lastname{Mackenzie}\inst{5}
}
%----------------------------------------------------------------------------
\institute{%
Physics Department, Boston University, Boston, Massachusetts 02215, USA
 \and
Physics Department, Columbia University, New York, New York 10027, USA
 \and
Department of Physics and Astronomy, University of Utah, Salt Lake City, Utah 84112, USA
 \and
Thomas Jefferson National Accelerator Facility, Newport News, Virginia 23529, USA
 \and
Fermi National Accelerator Laboratory, Batavia, Illinois 60510, USA
}
%----------------------------------------------------------------------------
\abstract{% 
  In October, 2016, the US Department of Energy launched the
  Exascale Computing Project, which aims to deploy exascale computing
  resources for science and engineering in the early 2020's. The
  project brings together application teams, software developers, and
  hardware vendors in order to realize this goal. Lattice QCD is one
  of the applications. Members of the US lattice gauge theory
  community with significant collaborators abroad are developing
  algorithms and software for exascale lattice QCD calculations. We
  give a short description of the project, our activities, and our
  plans.  }
%----------------------------------------------------------------------------
\maketitle
%----------------------------------------------------------------------------
\section{Introduction}\label{sec:intro}

The Exascale Computing Project (ECP) is a joint undertaking by the US
Department of Energy Office of Science (DOE-SC) and the US
National Nuclear Security Administration (NNSA) \cite{ECP}.  Launched in
October, 2016, its goal is to develop computers, software, and
algorithms that will provide fifty times the power of today’s fastest
machines.  The initial ``advanced architecture” is planned for the year
2021.  At least one, possibly two computers are envisioned.  A
``capable exascale” machine, based on ECP research and development, is
planned for 2023. The procurement process is not part of the ECP.

The US is not alone in advancing to the exascale. There are similar
initiatives in Europe \cite{ECexascale}, Japan \cite{JPpostK}, and
China \cite{ChinaSunway}, some of them already well underway.

Exascale computing presents a host of technical hardware challenges.
Here are some of the most important ones:
\begin{enumerate}
\item Increased parallelism.  We expect a thousandfold increase in parallelism
  over that of today's systems.
\item Memory, network, and storage efficiencies.  These must be improved
  in a manner consistent with increased computational rates and data-movement
  requirements.
\item Reliability. As systems grow in size and complexity, the ability 
  to adapt and recover from faults will become crucial.
\item Energy consumption. Electrical power consumption must be
  significantly reduced from today's standards.
\end{enumerate}
To meet these challenges will require considerable industrial innovation.

How is exascale capability defined?  It is not defined in terms of
a peak flop rate.  It must be meaningful to the mix of applications
that are expected to need computing at that scale.  It is defined as follows:
\begin{enumerate}
\item Fifty-fold increase in application performance.  With a facility on the
  scale of a ``leadership computing facility'' (such as the Argonne
  and Oak Ridge Leadership Computing Facilities) it should deliver 50
  times the performance of today’s 20 Petaflop systems for a wide
  range of applications.  This criterion is spelled out further below.
\item Power consumption.  The targeted range is 20–30 MW.
\item Resilience. The perceived fault rate should be less than one per week.
\item Software. It should includes a software stack that supports 
      a broad spectrum of applications and workloads.
\end{enumerate}

\section{National organization of the ECP}\label{sec:USorg}

The ECP is headed by Paul Messina (Argonne Leadership Computing
Facility)\footnote{Douglas Kothe assumed this leadership position on
October 1, 2017.} and managed through Oak Ridge National Laboratory.
The ECP partners with six major US DOE national laboratories, all with
strong high-performance computing expertise.  They are Argonne,
Lawrence Livermore, Lawrence Berkeley, Los Alamos, Oak Ridge, and
Sandia National Laboratories. Universities throughout the US are also
involved, as are industrial partners.

\begin{figure}
\centering
\includegraphics[width=12cm,clip]{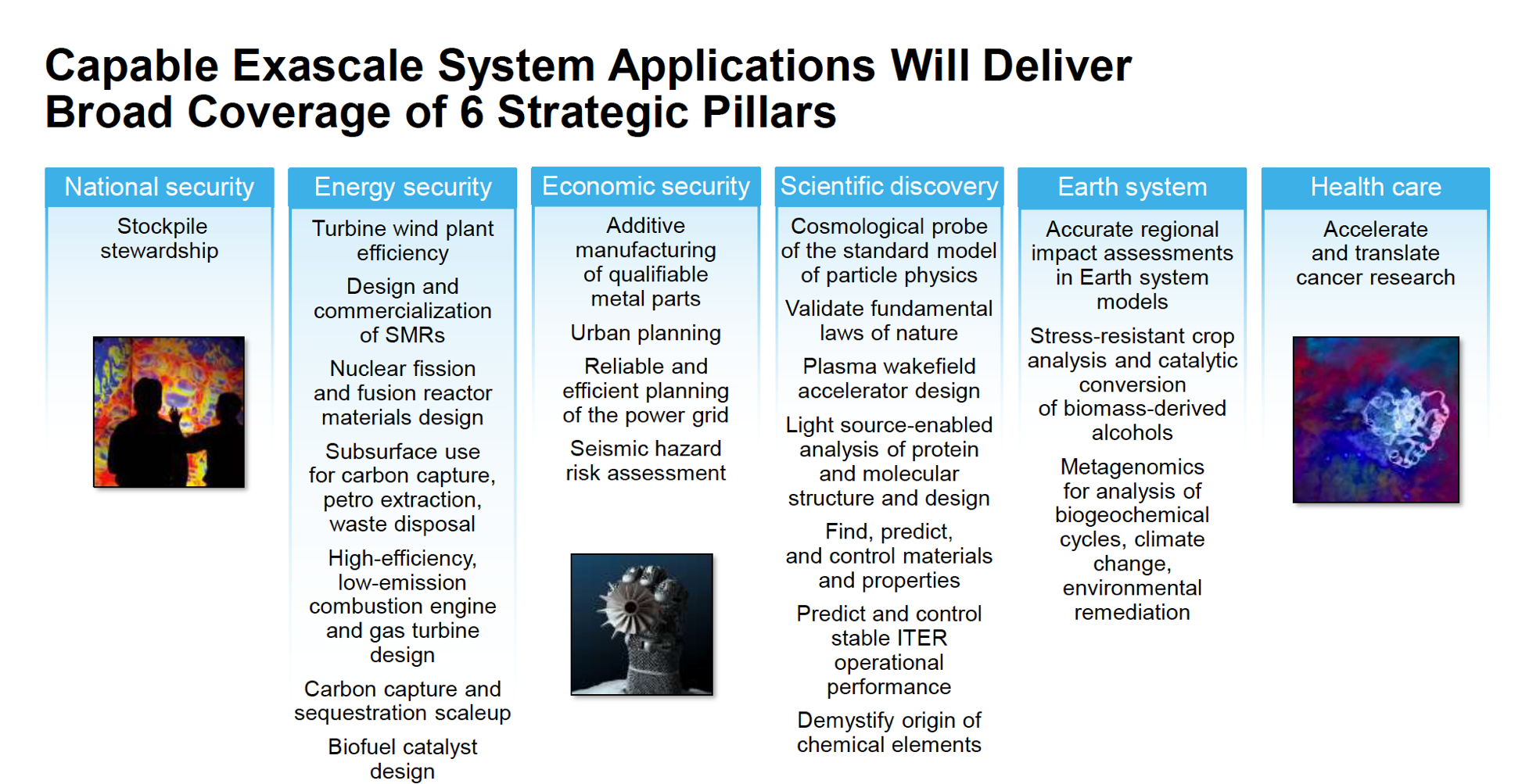}  
  \caption{Illustration of the range of applications targeted by the
    ECP from \cite{Messina}.}
  \label{fig:pillars}
\end{figure}

The ECP development effort follows a ``holistic co-design'' principle
that integrates the efforts of application developers and the
developers of software and hardware technology.
Figure~\ref{fig:pillars} gives an idea of the mix.  There are some
thirty application development projects involved.  They were selected
to represent a diverse mix of the HPC scientific and ``mission''
programs supported by the DOE.  They include applications ranging from
cosmology to electrical power grid design.  The software technology
component includes some 80 teams with subjects ranging from MPI to
solvers.  Some thirty commercial vendors are involved in developing
hardware technology.

In the past year, the DOE announced its selection of six major vendors
under the ``PathForward'' program.  This is a US\$430 M joint
industry/DOE undertaking with US\$258 M being provided by the DOE in
stages and the balance provided by the industrial partners.  The
partners are AMD, Cray, Hewlett Packard Enterprise (HPE), IBM, Intel,
and NVIDIA.  As they proceed with hardware development, they will be
testing systems on testbeds being set up at the national laboratories.

\section{US Lattice QCD Exascale Development}\label{sec:LQCDorg}

The scientific goals for US exascale computing are shared by many in
the world community.  They include
\begin{itemize}
\item Assisting in the search for physics beyond the Standard Model by
  comparing precise calculations with precise experimental results.
  \begin{itemize}
  \item $B$ physics at lattice spacing smaller than $1/m_b$.
  \item Understanding CP violation in the K system
  \item Simulating ``non-standard'' weak K decays.
  \item A ten-fold reduction in uncertainties in weak matrix elements.
  \end{itemize}
\item Calculations at physical quark masses with QED.
\item Precision study of the structure and interactions of light nuclei
\end{itemize}

Some fifty lattice physicists participate in the lattice QCD
application development project.  Several postdocs and lab staff
receive some salary support, but the rest do not receive salary
support.  In addition to US lattice gauge theorists from labs and
universities, this effort includes significant participants at the
University of Edinburgh and NVIDIA corporation, and it includes applied
mathematicians and computer scientists.  The lattice effort is
coordinated through a steering committee with Paul Mackenzie as PI,
and four working groups:
\begin{itemize}
\item Solver research (Richard Brower)
\item Critical slowing down research (Norman Christ)
\item Computations of Euclidean correlation functions (Robert Edwards)
\item Software development (Carleton DeTar)
\end{itemize}

\paragraph{Solver research}
The research program investigates algorithms that will improve the
efficiency of the solvers used in exascale calculations.  An important
activity is to select or develop robust algorithms that reduce or
avoid internode communication.  Another is the application of
multigrid algorithms to the staggered fermion formulation
\cite{EWeinberg}.  Other topics incude combining multigrid algorithms
with deflation, domain decomposition solvers, and
all-mode-averaging with eigenvector compression \cite{CLehner}.

\paragraph{Critical slowing down} 
Exascale will make it possible to do calculations at a fine enough
lattice spacing that critical slowing down of gauge-field generation
algorithms becomes an increasing concern.  While the most dramatic
consequence of this critical slowing down is the freezing of topology,
one expects the autocorrelation times of physical observables to grow
as the inverse square of the lattice spacing, $1/a^2$.  Even with open
boundary conditions, the local diffusion of topology from the
boundaries into the bulk is expected to slow as $1/a^2$

\paragraph{Algorithms for Wick contractions}
For lattice QCD calculations of hadronic matrix elements of even light
nuclei, dealing with all the quark-line tensor (Wick) contractions
presents a considerable combinatoric problem.  So it is important to
develop efficient strategies that reduce the cost of calculation.
First steps to reducing the cost involve finding an optimal set of
``edge reductions'' that maximize reuse of building blocks, and to
rearrange the graph order to lower the number of temporaries.

\paragraph{Software development}
For many years the USQCD community has been developing
lattice-QCD-specific community codes, including the DOE SciDAC suite:
QDP/C, QDP++, and QLUA \cite{SciDAC}.  Specifically for GPUs, QUDA
provides excellent performance\cite{QUDA}.  The several US lattice
collaborations have also developed code bases more specific to the
scientific goals of the respective collaborations, such as Chroma
\cite{Chroma,Edwards:2004sx}, the CPS code \cite{CPS}, and the MILC
code \cite{MILC}.

Drawing from this experience, we are formulating a new data parallel
applications programming interface (API) that supports the algorithms
under development for lattice QCD.  These include support for
multigrid and domain-decomposed solvers.  The API design should be
extensible and allow for rapid prototyping of new algorithms.

We are also looking for implementations that port readily between many
core and GPU architectures and yet deliver excellent performance on
both.  We are particularly interested in the Grid software system
being developed by Peter Boyle and collaborators at Edinburgh
\cite{Boyle:2016lbp,Grid}.  It was originally designed to perform
especially well on the vectorized many-core Intel KNL architecture.
However, its design is flexible, so we have begun experimenting with
strategies for porting it to GPUs \cite{Meifeng}.

In preparation for formulating a data-parallel API, the software group
has laid out the required functionality and assessed semantic
alternatives for expressing it.  We are looking at both C++-11 and Nim
\cite{NIM}.  We have been using the Grid software system as an example
of C++ semantics and the QEX framework of James Osborn and
collaborators for Nim \cite{Jin:2016ioq,QEX}.  We give a few examples
of each.

Parallel transport is obviously important for QCD calculations.
In Grid the syntax is
\begin{verbatim}
tmp = CovShiftForward(U[mu], mu, src);
\end{verbatim}
where both \verb|tmp| and \verb|src| are lattice fields with a color index.
In QEX the syntax is
\begin{verbatim}
let T = newTransporters(U, src, 1)
tmp = T[mu] ^* src
\end{verbatim}
For another example, for multigrid algorithms we start from a fine
grid, subdivide it into blocks, and project to a coarse grid, using a
set of $n$ basis vectors as block projectors.  In Grid the syntax for this
operation is
\begin{verbatim}
blockProject( coarseData, fineData, Basis );
\end{verbatim}
and in QEX it is
\begin{verbatim}
let R = newRestrictor(coarseLattice, Basis)
coarseData := R ^* fineData
\end{verbatim}
Data-parallel operations hide loops over all sites.  However, it is
often desirable to fuse multiple loops into a single loop to avoid
creating temporary intermediate fields and wasting time with avoidable
memory traffic.  One way loop fusion can occur in data-parallel
operation is through the use of expression templates.  Consider the operation
\begin{verbatim}
  q = a*x + y
  r = y + b*q;
\end{verbatim}
where \verb|q|, \verb|r|, \verb|x|, and \verb|y| are lattice fields
and \verb|a| and \verb|b| are scalar constants.  With expression
templates, the first operation can be compiled into a single, threaded
site loop.  But the second operation gets a separate loop, an \verb|y|
must be kept as a temporary intermediate.  If we are willing to expose
the site index, we can fuse the loops.  Here is an example from Grid:
\begin{verbatim}
PARALLEL_FOR_LOOP
for(s = 0; s < nsites; s++){
  q[s] = a*x[s] + y[s];
  r[s] = y[s] + b*q[s];
}
\end{verbatim}
where the \verb|PARALLEL\_FOR\_LOOP| generates an OpneMP pragma, requesting the
the fused loop be threaed.  With QEX it is proposed to do it this way:
\begin{verbatim}
fuse:
  q = a*x + y
  r = y + b*q
\end{verbatim}
So with QEX, an explicit directive is provided to force fusion, but
the site index is not exposed.

The US ECP project could make an interesting case study for a
sociologist.  The management regimen is quite a bit stricter and more
corporation-like than most of us outside industry are used to.  We are
required to use sophisticated project management systems, namely,
Atlassian's Confluence and Jira to manage our work, defining and
setting intermediate goals and deliverables to keep the work on track.
Clearly, too much such management risks hampering creativity, while
too little risks failing to reach goals.

Integrating the work of many projects is also presents a sociological
challenge.  ``Match making'' was the central purpose of the first
annual plenary ECP meeting, which took place in Knoxville, Tennessee,
January, 2017.  For Lattice QCD, the strongest interest thus far is in
the Kokkos \cite{Kokkos} and Trilinos \cite{Trilinos} software efforts,
OpenMP, OpenACC, and MPI design, HDF5, and checkpointing schemes.

Success of the entire ECP project will be measured by the ability of
most of the included applications to reach the goal of a fifty-fold
increase in computing capability.  All project improvements are to be
included in determining success, namely, hardware, systems design,
software, and algorithms.  To measure improvement in performance, each
application is required to specify a set of application-specific ``Key
Performance Parameters (KPP)'' and ``Figures of Merit (FOM)''.  For
Lattice QCD, we have defined a set of lattice-generation and
physics-analysis tasks and measured the time it takes to complete
those tasks on a specified fraction of a leadership-class facility,
such as those at Argonne or Oak Ridge National Laboratories.  We
are then aiming for a fifty-fold improvement in the time required to
complete the same tasks on the eventual exascale replacement of the same
facility.

\section{Conclusion and Outlook}\label{sec:concl}

The US DOE Exascale Computing Project is helping to accelerate the
arrival of wide-spread exascale computing.  Coordination among
hardware vendors, systems development teams, and application
developers is crucial to its success. We hope our efforts contribute
to the success of the lattice QCD program in the US and to the
international effort as well.

\section{Acknowledgment}

This research was supported by the Exascale Computing Project
(17-SC-20-SC), a collaborative effort of the U.S. Department of Energy
Office of Science and the National Nuclear Security Administration.

\clearpage
\bibliography{Lattice2017_396_DeTar}

%%%%%%%%%%%%%%%%%%%%%%%%%%%%%%%%%%%%%%%%%%%%%%%%%%%%%%%%%%%%%%%%%%%%%%%%%%%%%
\end{document}